\documentclass[aps,prb,superscriptaddress,showpacs,citeautoscript,twocolumn]{revtex4}

\usepackage{graphicx}
\usepackage{amsmath}
\usepackage{amssymb}

\newcommand{\bea}{\begin{eqnarray*}}
\newcommand{\eea}{\end{eqnarray*}}
\newcommand{\bne}{\begin{equation*}}
\newcommand{\ede}{\end{equation*}}

\newcommand{\bnen}{\begin{equation}}
\newcommand{\eden}{\end{equation}}
\newcommand{\bean}{\begin{eqnarray}}
\newcommand{\eean}{\end{eqnarray}}
\newcommand{\bnsn}{\begin{subequations}}
\newcommand{\edsn}{\end{subequations}}

\newcommand{\bna}{\begin{array}}
\newcommand{\eda}{\end{array}}

\newcommand{\f}{\frac}

\newcommand{\igr}[2][]{\includegraphics[#1]{#2}}
\newcommand{\spinor}[2]{\left[\bna{c} #1 \\[1.5ex] #2 \eda\right]}
\newcommand{\Z}{\mathbb{Z}}
\renewcommand{\i}{\mathbf{i}}
\renewcommand{\j}{\mathbf{j}}
\renewcommand{\k}{\mathbf{k}}
\newcommand{\h}{\mathbf{h}}

\newcommand{\ki}{k_p}
\newcommand{\ko}{k_n}
\newcommand{\psii}{\psi^{(p)}}
\newcommand{\psio}{\psi^{(n)}}

\renewcommand{\vec}[1]{\text{\boldmath{$ #1 $}}}
\begin{document}
\title{Electron Flow in Circular n-p Junctions of Bilayer Graphene}

\author{Cs. P\'eterfalvi}
\affiliation{Department of Physics of Complex Systems,
E\"otv\"os University,
H-1117 Budapest, P\'azm\'any P\'eter s\'et\'any 1/A, Hungary}

\author{A. P\'alyi}
\affiliation{Department of Physics of Complex Systems,
E\"otv\"os University,
H-1117 Budapest, P\'azm\'any P\'eter s\'et\'any 1/A, Hungary}
\affiliation{Department of Physics, University of Konstanz, D-78457 Konstanz, Germany}

\author{J. Cserti}
\affiliation{Department of Physics of Complex Systems,
E\"otv\"os University,
H-1117 Budapest, P\'azm\'any P\'eter s\'et\'any 1/A, Hungary}

\date{\today}

\begin{abstract}

We present a theoretical study of electron wave functions in ballistic circular $n$-$p$ junctions
of bilayer graphene.
Similarly to the case of a circular $n$-$p$ junction of monolayer graphene, we find that
(i) the wave functions form caustics inside the circular region, and
(ii) the shape of these caustics are well described by a geometrical
optics model using the concept of a negative refractive index.
In contrast to the monolayer case, we show that the strong focusing effect
is absent in the bilayer.
We explain these findings in terms of the angular dependence of Klein tunneling
at a planar $n$-$p$ junction.

\end{abstract}

\maketitle


\section{Introduction}

The interface of an $n$-$p$ junction (NPJ) of graphene\cite{Novoselov-fieldeffect,novoselov-2ddirac,Neto2009}
is fully transparent for electrons approaching it with
a perpendicular incidence\cite{cheianov-veselago, cheianov-smooth, katsnelson-klein, Beenakker2009,Zhang-gnpj,Fogler-disorderedgnpj}.
Electrons approaching the interface at a finite angle are still
transmitted with a high probability provided that
the transition between the $n$ and the $p$ regions
is sharp enough\cite{cheianov-smooth}.
As proposed recently by Cheianov \emph{et al.}\cite{cheianov-veselago}, this high
transparency of the interface offers a way to use
the graphene NPJ as an electronic lens.
The refraction of electron rays in this
system follows Snell's law with a negative
refractive index, which is a consequence of the
fact that the wave vector and the velocity of the
valence band quasiparticles in the $p$ region are
antiparallel.
In the case of a point-like source of electrons
on the $n$ side of the interface,
the NPJ provides perfect focusing
of the emitted electrons on the $p$ side
if $k_n = k_p$, where $k_n$ ($k_p$) is the
wave number in the $n$ ($p$) region.
If $k_n \neq k_p$, the sharp focus transforms
into a smeared focus and a pair of caustics.
(For a review on the theory and classification of
caustics see Ref. \onlinecite{Berry-catastrophe-optics}.)

According to our earlier theoretical analysis\cite{cserti-caustics},
focusing and caustic formation also arises
in \emph{circular} $n$-$p$ junctions of graphene, where
the $n$ ($p$) region is defined as the area
outside (inside) a circle.
Such a device is found to be able to focus an incident
parallel beam of electrons into a certain spot inside the
$p$ region, however the focusing is imperfect and caustic
formation arises even if $k_n = k_p$.

The interband or Klein
tunneling\cite{katsnelson-klein,Beenakker2009} of carriers in bilayer
graphene\cite{Novoselov-bilayer,McCann-bilayerreview}
is remarkably different from the same process in monolayer graphene.
Namely, the bilayer NPJ \emph{reflects} normally
incident electrons with unit probability\cite{katsnelson-klein}.
This difference leads to the anticipation that the patterns
of electron flow in planar or circular bilayer graphene $n$-$p$
junctions are distinct from the patterns in their monolayer
counterparts.

In this work, we investigate the possibility of controlling
the electron flow in bilayer graphene by using a gate-defined
circular NPJ.
We provide an exact solution of the effective Schr\"odinger
equation in the presence of a step-like circular potential barrier.
Using the exact wave functions we demonstrate that in contrast to the monolayer case,
the focusing of a parallel electron beam is not possible
in the circular bilayer NPJ.
However, we find that caustic formation remains a sizeable and
possibly observable effect even in the bilayer.
We also calculate the angular dependence of
transmission probability in a planar NPJ, and use the results
of this calculation to interpret the
absence of focusing and the presence of caustic formation
in circular junctions.

The paper is organized as follows. In Section \ref{sec:circular}
we solve the effective Schr\"odinger equation
modelling the circular NPJ in bilayer graphene using
the method of partial waves.
In Section \ref{sec:planar} we calculate the angular dependence
of transmission probability in a planar NPJ, and discuss
the results of Section \ref{sec:circular} in terms of the
transmission probability function.
In Section \ref{sec:discussion} we discuss the validity of the
model we use, give a brief overview of related experiments,
and provide a short conclusion.

\section{Electron flow in a circular $n$-$p$ junction}
\label{sec:circular}

\begin{figure}[hbt]
\igr[scale=0.45]{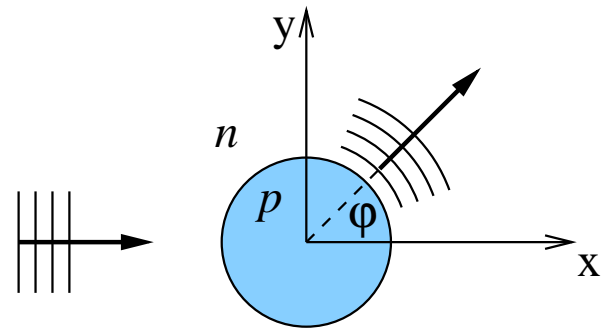}
\caption{\label{setup:fig}
(Color online)
An incident plane wave of electrons in bilayer graphene is scattered by a
circular $n$-$p$ junction created by a gate-induced circular potential barrier $V(r)$.
In the $n$ ($p$) region the Fermi energy lies in the conduction (valence) band.}
\end{figure}

In this Section we consider the scattering of an electron plane
wave on a circular $n$-$p$ junction in bilayer graphene (see Fig.\ref{setup:fig}).
Our goal is to calculate the exact scattering wave function
as a function of system parameters
and to identify the characteristics of the flow of
electrons \emph{inside} the circular $p$ region.
We will focus on the case when the radius of the circle is much
larger then the electron wavelength, since in that regime we
expect a correspondence between results of
the quantum mechanical model and
a simplified description based on principles of geometrical optics.

To model the two-dimensional electron flow in bilayer graphene,
we use a two-component envelope function Hamiltonian
which has been derived by
McCann and Fal'ko\cite{McCann-bilayerprl}.
The derivation of this Hamiltonian starts
from a simple tight-binding model of bilayer graphene,
which contains only nearest-neighbor intralayer and interlayer
hopping matrix elements.
These hopping matrix elements are usually denoted\cite{Neto2009,McCann-bilayerreview} by
$\gamma_0$ and $\gamma_1$, respectively.
This tight binding model predicts that the
valence and conduction bands of bilayer graphene
are touching at the $K$ and $K'$ points,
i. e. at the two nonequivalent corners of the hexagonal Brillouin zone.
The Fermi energy of undoped bilayer graphene lies exactly
at the energy corresponding to these touching points.
In the vicinity of the $K$ point, the dispersion relations
describing the conduction and valence bands
are both quadratic, $E_\pm (\vec k) \approx \pm \hbar^2 (\vec k-\vec K)^2 / 2m$,
where $m = \frac{2\hbar^2\gamma_1}{3a^2\gamma_0^2} \approx 0.054 m_0$
is the effective mass.
Here the $+$ ($-$) sign refers to the conduction (valence) band,
$a$ is the lattice constant and $m_0$ is the free electron mass.
In this low-energy regime the quasiparticle wave functions
can be characterized by the eigenfunctions of the $2\times 2$
effective Hamiltonian
\bnen
H_0 = -\frac{1}{2 m}\left[ \begin {array}{cc} 0&p_-^2\\\noalign{\medskip}p_+^2&0\end {array} \right],
\eden
where $p_{\pm} = (p_x \pm ip_y)$.
Note that this Hamiltonian describes the valence
band and conduction band states simultaneously.
Similar statements are true for the vicinity of the $K'$ point.
We note that the dispersion relation and the effective Hamiltonian
becomes more complex if one includes second-nearest-neighbor
interlayer hopping matrix elements in the tight-binding model.
In particular, such terms lead to a trigonal warping\cite{McCann-bilayerreview,McCann-bilayerprl,Cserti-trigonalwarping}
of the quasiparticle dispersion.
We comment on the significance of trigonal warping in
Section \ref{sec:discussion}.

We model the gate-defined circular potential barrier by a step-like
potential $V(r) = V_0 \Theta(R-r)$, where $\Theta$ denotes
the Heaviside function.
Hence the complete Hamiltonian of the system under study is
\bnen
\label{eq:hamiltonian}
H = H_0 + V(r)  \openone,
\eden
where $\openone$ is the $2\times2$ unit matrix.
The validity of this model will be discussed in Section \ref{sec:discussion}.
Note that the same model was used recently to calculate the
lifetime of quasibound states in a similar system\cite{Matulis-bilayerdot}.

We concentrate on the regime where the potential barrier forms an $n$-$p$
junction, i.e. the Fermi energy $E_F$ of the electrons
lies between the Dirac point of the bulk and the top of the
potential barrier ($0<E_F<V_0$).
In this case, the region outside the circle of radius $R$ contains
electrons in the conduction band ($n$-type), whereas
the region inside the circle contains holes in the valence band
($p$-type).

Our aim is to consider the scattering of an incident electron plane
wave coming from the $n$ region along the $x$-axis as shown in Fig. \ref{setup:fig} and having energy $E_F$.
Such an electron has the following two-component wave function\cite{katsnelson-klein}:
\bnen
\label{eq:planewave}
\phi(x,y) = e^{i\ko x}\frac 1 {\sqrt 2} \spinor{1}{-1},
\eden
where $\ko =\sqrt{2mE_F}/\hbar$.
In order to derive the wave function describing the scattering
of this plane wave, we first treat the scattering
of cylindrical waves and then utilize the fact that
the plane wave $\phi(x,y)$ is a certain linear combination of
cylindrical waves.
Here we note that scattering theory has been used recently to predict
transport properties of disordered\cite{Katsnelson-bilayer,Kechedzhi,Adam-bilayer,
Culcer-bilayer} and ballistic\cite{Braun-chirality} bilayer graphene
structures.

The system has a circular symmetry around the origin,
hence the Hamiltonian commutes with a ´pseudo angular momentum' operator
$J_z = -i\hbar \partial_\varphi+\hbar \sigma_z$,
where $\partial_\varphi$ is the derivative with respect to
the angular polar coordinate and $\sigma_z$ is the
third Pauli matrix.
The presence of this symmetry simplifies the forthcoming
calculations.

Using the properties of Bessel functions\cite{abramowitz}
it can be shown that in the $n$ region, for any
integer $j$ the wave functions
\bnsn \label{eq:nregion}
 \bean
   \h_{j}^{(1)}(r,\varphi) &=& \spinor{H^{(1)}_{j-1}(\ko r) e^{-i \varphi}}{H^{(1)}_{j+1}(\ko r) e^{i \varphi}} {e^{i j \varphi}}, \\
   \h_{j}^{(2)}(r,\varphi) &=& \spinor{H^{(2)}_{j-1}(\ko r) e^{-i \varphi}}{H^{(2)}_{j+1}(\ko r) e^{i \varphi}} {e^{i j \varphi}}, \\
   \k_{j}(r,\varphi) &=& \spinor{K_{j-1}(\ko r) e^{-i \varphi}}{K_{j+1}(\ko r) e^{i \varphi}} {e^{i j \varphi}}
 \eean
\edsn
are simultaneous eigenfunctions of $H$ and $J_z$, with
eigenvalues $E_F$ and $\hbar j$, respectively.
We denote the radial polar coordinate with $r$.
Here $H_m^{(1)}$, $H_m^{(2)}$ and $K_m$ denote
Hankel functions of first and second kind and
the modified Bessel function which is bounded for
large arguments\cite{abramowitz}, respectively.
There exists a solution similar to those in
Eq. \eqref{eq:nregion}, containing the modified Bessel
function $I_m$.
We disregard it because $I_m$ diverges for large arguments.
Analysis of the quantum mechanical current density
in state $\h_j^{(1)}$ ($\h_j^{(2)}$) shows that
it is an outgoing (incoming) cylindrical wave.
On the other hand, $\k_j$ is an evanescent cylindrical wave
which does not carry current in the radial direction.

Inside the circular $p$ region, the regular eigenfunctions
of the Hamiltonian $H$ having energy $E_F$ are
\bnsn \label{eq:pregion}
 \bean
   \j_{j}(r,\varphi) &=& \spinor{J_{j-1}(\ki r) e^{-i \varphi}}{-J_{j+1}(\ki r) e^{i \varphi}} {e^{i j \varphi}}, \\
   \i_{j}(r,\varphi) &=& \spinor{I_{j-1}(\ki r) e^{-i \varphi}}{-I_{j+1}(\ki r) e^{i \varphi}} {e^{i j \varphi}}.
 \eean
\edsn
Here $\ki = \sqrt{2m(V_0-E_F)}/\hbar$ and $j$ is an
arbitrary integer.
Similarly to the wave functions in the $n$ region,
$\j_j$ and $\i_j$ are eigenfunctions of $J_z$ with
an eigenvalue $\hbar j$.
We disregard other eigenfunctions of $H$ which
are divergent at the origin.

Now we consider the scattering of a single incoming
cylindrical wave, $\h^{(2)}_j$.
Since $[H,J_z]=0$, the pseudo angular momentum
does not change during the scattering process, therefore
the complete wave function describing the scattering
can be written as
\bnsn \label{eq:cylindrical_ansatz}
 \bean
   \psio_j &=& \h^{(2)}_j+ S_j \h^{(1)}_j + A_j \k_j,  \\
   \psii_j &=& B_j \j_j + C_j \i_j,
 \eean
\edsn
in the $n$ and $p$ regions, respectively.
The coefficients $S_j$, $A_j$, $B_j$ and $C_j$
have to be determined from the boundary conditions
at the interface of the NPJ:
the wave functions and their derivatives have
to be continuous at $r=R$.
Due to the two-component nature of the
wavefunctions, the two boundary conditions
result in an inhomogeneous linear system with four equations
and the four coefficients as unknowns.
This system can be solved analytically.

Having the coefficients $S_j$, $A_j$,
$B_j$ and $C_j$ in hand, one can determine the
wave function describing the scattering of
the plane wave $\phi$ in Eq. \eqref{eq:planewave}.
Making use of the fact that\cite{abramowitz}
\bnen
\label{eq:partial1}
e^{i k x} = \sum_{m \in \Z} i^m J_m(k r) e^{im \varphi},
\eden
it can be shown that the plane wave $\phi$
can be written as a linear combination of incoming
and outgoing cylindrical waves:
\bnen
\phi = \frac{1}{i \sqrt 8} \sum_{j \in \Z}
i^j \left(
\h^{(1)}_j+\h^{(2)}_j
\right).
\eden
This expansion allows us to use the coefficients determined
from the analysis of partial waves to derive the
wave function describing the scattering of the plane wave.
In the $n$ region,
\bnen
\psi^{(n)} = \phi + \frac{1}{i\sqrt{8}}\sum_{j\in\Z} i^j
\left[
(S_j-1) \h^{(1)}_j + A_j \k_j
\right],
\eden
and in the $p$ region
\bnen
\psi^{(p)} = \frac{1}{i\sqrt{8}}\sum_{j\in\Z} i^j
\left(
B_j \j_j + C_j \i_j
\right).
\eden
The complete wave function $\psi$ is constructed
by tailoring $\psi^{(n)}$ and $\psi^{(p)}$.
It is built up from cylindrical waves having energy $E_F$,
therefore $\psi$ is also an energy eigenstate with energy
$E_F$.
Since the cylindrical waves fulfill the boundary conditions
at $R$, $\psi$ also fulfills them.
Finally, since in the $n$ region $\psi$ contains only the
plane wave and outgoing and evanescent cylindrical waves
(no incoming wave), we conclude that $\psi$ is the wave
function which describes the scattering of the incident
plane wave.

\begin{figure}[hbt]
\igr[width=86mm]{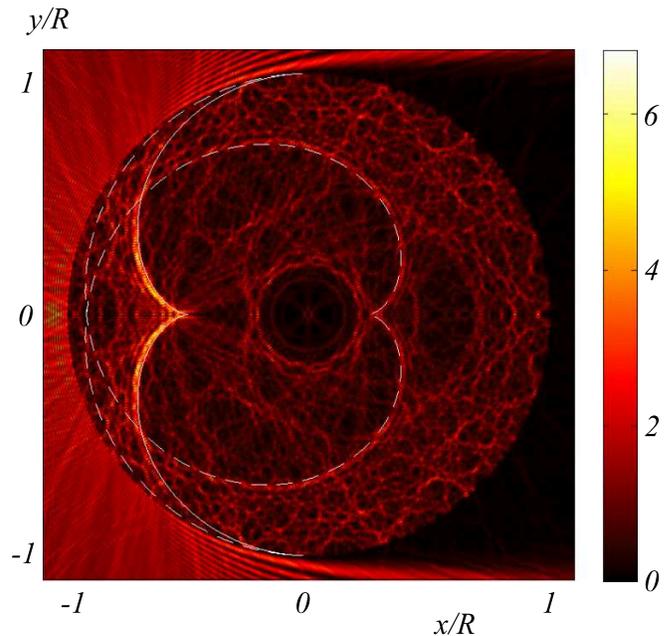}
\caption{\label{wavefn-1:fig}
(Color online)
The spatial dependence of the intensity of the wave
function $|\psi(\vec r)|^2$ is plotted in the scattering area.
Here $\ko R=300$, and $\ki R=300$ corresponding to $n = -1$.
The solid (dashed) line corresponds to the caustic for $p=1$ ($p=2$), where $p$ denotes the number of chords inside the NPJ \cite{cserti-caustics}.}
\end{figure}
\begin{figure}[hbt]
\igr[width=86mm]{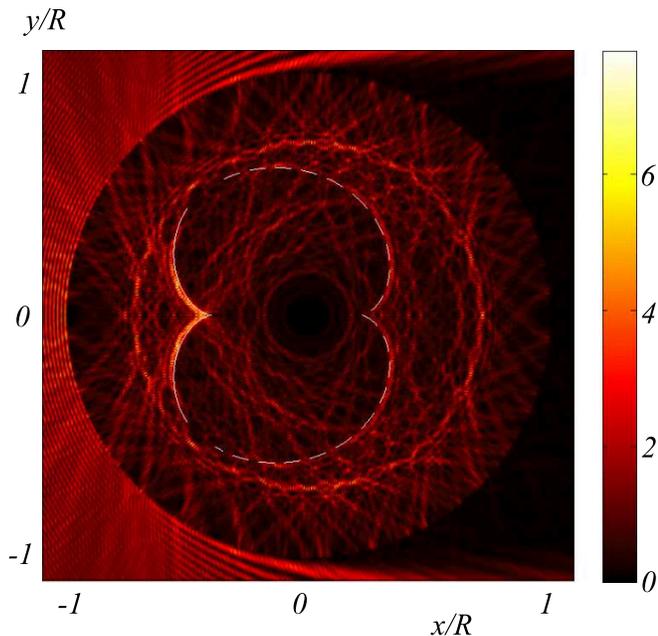}
\caption{\label{wavefn-2:fig}
(Color online)
The same as in Fig.~\ref{wavefn-1:fig} with
$\ko R= 200$ and $\ki R= 300$ corresponding to $n= -1.5$.}
\end{figure}

Numerical results for the spatial dependence of the
magnitude of the complete scattering state
($|\psi(\vec r)|^2$) are shown in
Fig. \ref{wavefn-1:fig} and Fig. \ref{wavefn-2:fig}
for two different set of parameters.
In both cases, well-defined patterns of the
electron flow can be identified,
the wave function magnitude is sharply peaked
close to the solid curve.
This effect is almost identical to the one predicted
for circular NPJs of monolayer graphene\cite{cserti-caustics}.
As we will argue in Section \ref{sec:planar},
the geometrical optics model developed in Ref. \onlinecite{cheianov-veselago}
and Ref. \onlinecite{cserti-caustics} for single layer graphene is
also applicable for bilayer with certain restrictions.
According to the referred theories, the refraction of the incident
electrons is governed by Snell's law with a negative refractive index.
After the refraction, the electrons enter the $p$ region of the
junction, and the envelope of the electron rays form a caustic.
These caustics can be identified in the quantum mechanical
charge density, as it is revealed by Figs. \ref{wavefn-1:fig}
and \ref{wavefn-2:fig}.

Despite the apparent similarities of the monolayer
and bilayer case, the analogy is not complete.
In a circular monolayer NPJ the charge density
is maximal close to the meeting point of the
two caustic lines, which means that the
interface between the $n$ and $p$ regions
provides strong focusing of the incident electrons\cite{cserti-caustics}.
This feature is missing in our results for the
bilayer.
In Section \ref{sec:planar}
we will show that the absence of focusing is
connected to a general characteristic
of interband (Klein) tunneling in bilayer
graphene.

\section{Transmission in a planar $n$-$p$ junction}
\label{sec:planar}

In this section we study the refraction of electron plane waves at
a planar $n$-$p$ junction of bilayer graphene.
We derive the counterpart of Snell's law for this system,
and calculate how the probability of transmission depends on the
propagation direction of the incident electron.
The obtained results will be used to explain our findings for the
circular NPJ (Section \ref{sec:circular}).

The studied system consists of a sheet of bilayer graphene in the $x$-$y$ plane
which is $n$-type for $x<0$ and $p$-type for $x>0$.
The electrostatic potential which creates these regions is modelled
by a step-like function $V(x,y)=V_0\Theta(x)$.
We consider a conduction electron plane wave incident from the $n$ side of the
junction.
We assume that the propagation direction of the plane wave is given by the angle
$\alpha \in [-\pi/2,\pi/2]$, and it has energy $E_F$ ($0<E_F<V_0$).

To derive the Snell's law for planar $n$-$p$ junction of bilayer graphene
we follow Ref. \onlinecite{cheianov-veselago}.
The length of the wave vector in the $n$ ($p$) region is $\ko$ ($\ki$),
and the length of the corresponding group velocity is $v_n$ ($v_p$).
(We assume that the plane wave is refracted, and
do not consider the case of total reflection here.)
The incident electron has the velocity $v_n(\cos\alpha,\sin\alpha)$ and
wave vector $\ko(\cos\alpha, \sin\alpha)$.
At the interface this electron is partially reflected with velocity
$v_n(-\cos\alpha,\sin\alpha)$ and wave vector $\ko(-\cos\alpha,\sin\alpha)$.
We denote the direction of propagation of the refracted wave by $\beta$, hence
the velocity of the refracted wave is $v_p(\cos\beta,\sin\beta)$.
Since the refracted wave is in the valence band, its velocity is antiparallel
with its wave vector, and thus the corresponding wave vector is
$\ki(-\cos\beta,-\sin\beta)$.
The translational invariance of the system along the $y$ direction implies
that the $y$ component of the wave vector must not change during the refraction,
i.e. $\ko \sin\alpha=-\ki\sin\beta$, which results in Snell's law with
a \emph{negative} refractive index $n=-\ki/\ko$,
\bnen
\label{eq:snell}
\frac{\sin\alpha}{\sin\beta}=n<0.
\eden

This form of Snell's law is identical to the one found for monolayer graphene NPJs.
Consequently, the mathematical formula describing the caustic lines formed by
the electron rays in a circular bilayer NPJ is also identical to the one derived
for the monolayer case.
This formula is given for the monolayer in Eq. (9) of Ref. \onlinecite{cserti-caustics},
and it has been used to plot the solid curves in Figs.
\ref{wavefn-1:fig} and \ref{wavefn-2:fig}.
The correspondence between the description of the electron flow in terms of quantum mechanics and geometrical optics
is apparent from the figures:
the high-density regions of the quantum mechanical wave functions
are condensed in the vicinity of the caustic line.

We further investigate the refraction of electrons
at the $n$-$p$ interface by calculating the
probability of transmission as
the function of the angle of incidence $\alpha$.
The system is modelled by the Hamiltonian
$H=H_0+V_0\Theta(x)$.
The wave functions at the $n$ and $p$ regions
can be constructed using the results of Ref. \onlinecite{katsnelson-klein}.
In the $n$ region, the incident, reflected and evanescent modes are given
by
\bnsn
 \bean
    \psi_{\rm inc}(x,y)  &=&  e^{i k_{ny}y} e^{i k_{nx} x} \spinor{1}{-e^{2 i \alpha }},\\
    \psi_{\rm refl}(x,y) &=&  e^{i k_{ny}y} e^{-i k_{nx} x}\spinor{1}{-e^{-2 i \alpha }},\\
    \psi_{\rm ev,n}(x,y) &=&  e^{i k_{ny}y} e^{\kappa_n x} \spinor{1}{h(\alpha)},
 \eean
\edsn
where $k_{ny}=k_n\sin\alpha$, $k_{nx}=k_n\cos\alpha$, $\kappa_n=k_n\sqrt{1+\sin^2\alpha}$
and $h(\alpha)=(\sqrt{1+\sin^2\alpha}-\sin\alpha)^2$.
In the $p$ region, the refracted and the evanescent waves are
\bnsn
 \bean
   \psi_{\rm refr}(x,y) &=&  e^{i k_{ny} y} e^{-i k_{px} x} \spinor{1}{e^{2 i \beta}},\\
   \psi_{\rm ev,p}(x,y) &=&  e^{i k_{ny} y} e^{-\kappa_p x} \spinor{1}{-1/h(\beta+\pi)}.
 \eean
\edsn
Here $k_{px}=k_p\cos\beta$ and $\kappa_p=k_p\sqrt{1+\sin^2\beta}$.
Note that the refraction angle $\beta$ has to be determined from
Snell's law [Eq. \eqref{eq:snell}].

The wave function describing the reflection-refraction
process in the $n$ region is
$\psi_n=\psi_{\rm inc} + r\psi_{\rm refl}+a\psi_{\rm ev,n}$,
whereas in the $p$ region it is
$\psi_p=t\psi_{\rm refr} + b\psi_{\rm ev,p}$.
One has to determine the coefficients $r$, $a$, $t$ and $b$
from the boundary conditions which match the wavefunctions
and their derivatives at the interface $x=0$.
The transmission probability as a function of the
angle of incidence is
\bean
T(\alpha) &=& |t(\alpha)|^2 \f{\left<\psi_{\rm refr}|v_x|\psi_{\rm refr}\right>}{\left<\psi_{\rm inc}|v_x|\psi_{\rm inc}\right>} \mbox{, where}\\
v_x&=&\f{i}{\hbar}[H,x]=-\f{1}{m}\left[\bna{cc} 0 & p_- \\ p_+ & 0 \eda\right]
\eean
is the $x$-component of the current operator. Similarly, we found that $R(\alpha)=|r(\alpha)|^2=1-T(\alpha)$.

\begin{figure}
\igr[scale=0.8]{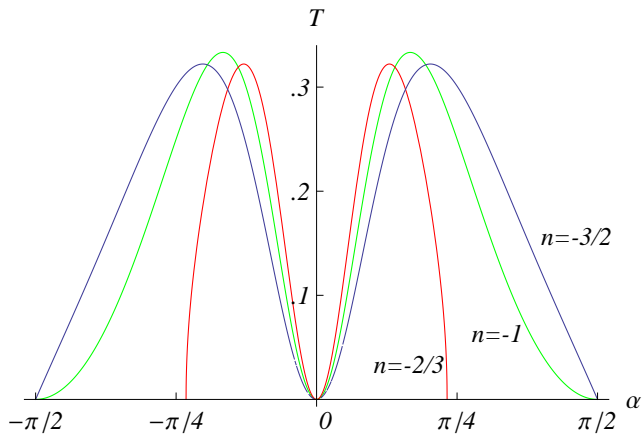}
\caption{\label{T:fig}
(Color online)
Angular dependence of transmission probability through a planar NPJ of bilayer
graphene.
The corresponding values of the refractive index $n$ are shown in the figure.}
\end{figure}

\begin{figure}[hbt]
\includegraphics[scale=0.7]{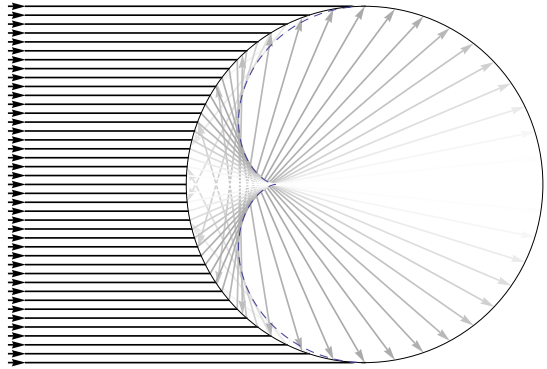}
\caption{\label{caustics_geo:fig}
(Color online)
Refraction of electron rays at the interface of
the $n$-$p$ junction.
The darkness of the refracted
rays reflects the transmission probability:
a white ray corresponds to $T=0$,
and a black ray would correspond to $T=1$.
The dashed curve shows the shape of the caustic
derived from Snell's law\cite{cserti-caustics}.
Only the rays with no internal reflections are displayed.
Here the refractive index is $n=-1$.
}
\end{figure}

In Fig. \ref{T:fig} we plot $T(\alpha)$ for three
different values of the refractive index.
A characteristic feature of all the three
cases is the absence of transmission for
perpendicular incidence, $T(0)=0$.
This behavior has been predicted and
explained with the chiral nature of the
quasiparticles of bilayer graphene\cite{katsnelson-klein}.
With the help of Fig. \ref{caustics_geo:fig} we argue that
the absence of transmission at
perpendicular incidence is responsible for the
complete suppression of focusing in the circular
junction we studied in Section \ref{sec:circular}.
Fig. \ref{caustics_geo:fig} shows several electron rays approaching the
circular $p$ region and being refracted at the $n$-$p$
interface.
An incoming electron ray can be characterized by
its impact factor $b$, which is the distance between
the incoming ray and the optical axis
(defined as the line containing the horizontal diameter of the circular $p$ region,
see Fig. 4 of Ref. \onlinecite{cserti-caustics}).
Note that the impact factor of rays entering the $p$ region is between $-R$ and $R$.
The angle of incidence, i.e. the angle between the incoming electron
ray and the local normal vector of the interface at the point of
incidence, can be calculated from simple trigonometry:
$\alpha(b) = {\rm arcsin} \frac{b}{R}$.
From this result one can express
(i) the propagation direction of the
refracted ray using Snell's law in Eq. \eqref{eq:snell}, and
(ii) the probability of transmission (=refraction) as a function of the impact
parameter $T(b)$, by combining the results for $T(\alpha)$ and $\alpha(b)$.
In Fig. \ref{caustics_geo:fig} the darkness of the refracted
electron rays reflects the transmission probability $T(b)$.
The figure indicates that the electron rays approaching
the junction in the close vicinity of the optical axis
does not have an appreciable probability of transmission at the
interface, which leads to the complete suppression of
the focusing effect (cf. Fig. 5 of Ref. \onlinecite{cserti-caustics}).

An other remarkable feature of the transmission functions
plotted in Fig. \ref{T:fig} is that
transmission is significant ($\geq 0.1$) in a wide
range between perpendicular ($\alpha=0$)
and nearly parallel ($|\alpha| \approx \pi/2$) incidence.
With respect to the circular junction shown in Fig. \ref{caustics_geo:fig},
it means that electron rays hitting the
NPJ further from the optical axis have a significant
probability to be refracted, and hence,
to form caustics along the dashed line of
Fig. \ref{caustics_geo:fig}.
This analysis built on the geometrical optics
approach and the transmission probability calculations
for the planar junction provides a qualitative explanation of the presence
of the well-defined wave function patterns
presented in Figs. \ref{wavefn-1:fig} and \ref{wavefn-2:fig}.

\section{Discussion and Summary}
\label{sec:discussion}

During the analysis of circular and planar NPJs,
we modelled the electrostatic potential as a
step-like function of position which does
not couple different valleys.
This assumption is justified if
$a\ll d\ll \lambda_n,\lambda_p$,
where $a$ is the lattice constant, $d$ is
the characteristic length describing the
width of the transition region between the $n$
and $p$ sides of the junction
and $\lambda_{n}$, $\lambda_p$ are the de Broglie wavelengths
of the considered quasiparticles in the $n$ and $p$ regions.
The first relation $a\ll d$ ensures the absence of intervalley
scattering at the interface, and
the second relation $d\ll \lambda_n,\lambda_p$ justifies the usage of a
step-like potential in our model.

Since the geometrical optics model is expected to capture the main
features of electron flow patterns only when the wavelength of the
electrons is much shorter then the size of the system,
the conditions $k_nR, k_pR\gg 1$ has to be fulfilled
to observe a pronounced caustic formation effect.

In the model Hamiltonian in Eq. \eqref{eq:hamiltonian} we neglected the
trigonal warping term\cite{McCann-bilayerprl,McCann-bilayerreview}
and used an approximate effective Hamiltonian which results
in a quadratic dispersion relation.
In bilayer graphene trigonal warping is strong only at very
low energies, when $E_F<\gamma_1(\gamma_3/\gamma_0)^2/4\approx1.15$meV.
Here $\gamma_0$, $\gamma_1$ and $\gamma_3$ are
hopping matrix elements of the standard tight-binding model
of bilayer graphene\cite{McCann-bilayerprl,McCann-bilayerreview},
and we used estimates for them from the review of
Castro Neto \emph{et al}.\cite{Neto2009}
For larger energies, the dispersion relation is
dominantly quadratic up to an energy
$E_F = \gamma_1/2 \approx 200$meV,
where it crosses over to a mostly linear dispersion.\cite{McCann-bilayerprl,McCann-bilayerreview}
Therefore our model should give a good description of
quasiparticles having energies between $1.15$meV and $200$meV.

To give a numerical example of parameters which
fulfill the above criteria,
we consider a bilayer graphene circular NPJ with
Fermi energy $E_F=10$meV, gate potential $V_0=20$meV,
transition region width $d=10$nm and
$p$ region radius $R=1\mu$m.
Then $k_n=k_p\approx 0.12$nm$^{-1}$,
$\lambda_n=\lambda_p\approx 53$nm,
$k_n R = k_p R \approx 118$ and the refractive index $n=-1$.
For these experimental parameters the model we used is expected to
give a good description of electron dynamics in the circular
bilayer NPJ, and the relation $k_n R = k_p R \approx 118 \gg 1$
is expected to ensure the strong caustic formation
effect in this system.

Finally, we summarize some experimental results which
support the feasibility of electron optics devices in
graphene in general.
Control of electron flow in a ballistic two-dimensional
electron system by means of gate-defined potential barriers
as refractive elements has been realized nearly two
decades ago\cite{Spector-electronfocusing}.
Direct imaging of the electron flow in a two-dimensional
electron system has also been carried out applying
scanning gate techniques\cite{topinka-1,topinka-2}.
Scanning tunneling microscopy has been used to show that
oscillations of the local density of states around a static impurity can be refocused
to a remote location\cite{Manoharan-qmirages}.
To realize similar experiments making use of the negative refraction index
in graphene, a trivial prerequisite is the
ability to fabricate gate-defined tunable NPJs, which has already
been reported by several
groups\cite{ozylmaz-exp,huard-exp,Liu-npjunctions-graphene,williams-exp,Gorbachev-pnpgraphene,Stander2009,Young-klein}.

In conclusion, we have carried out a theoretical
analysis of electron dynamics in circular
NPJs of bilayer graphene.
We demonstrated that such a system might be
used to control the flow of electrons,
similarly to previously realized and
proposed electron optics devices.
We have pointed out similarities and differences
between electron dynamics in the circular NPJ of bilayer and
monolayer graphene.
In both devices, electron paths form caustics inside
the circular $p$ region, and the form of these caustics can be
described with a geometrical optics model based on the concept
of negative refractive index.
The major difference is that the strong focusing of electrons,
which is a characteristic of the monolayer device, is completely
absent in the bilayer.
Our findings are explained in terms of the angular dependence of
transmission probability at a planar NPJ.

\acknowledgements
We gratefully acknowledge discussions with V.~Fal'ko. This work is supported by the Hungarian Science Foundation OTKA under
the contracts No. 48782 and 75529.


\bibliography{bilayer-focusing.bbl}

\end{document}